\documentclass[
superscriptaddress,
 amsmath,
 amssymb,
 aps,
pra,
floatfix,
twocolumn,
10pt
]{revtex4-2}
\pdfoutput=1
\usepackage[utf8]{inputenc}
\usepackage[T1]{fontenc}
\usepackage{babel}
\usepackage{graphicx}
\usepackage{amsfonts}
\usepackage{amssymb}
\usepackage{amsmath}
\usepackage{amsthm}
\usepackage{mathtools}
\usepackage{physics}
\usepackage[normalem]{ulem}
\usepackage{blindtext}
\usepackage{dsfont}

\usepackage{mathtools}
\usepackage{braket}

\renewcommand\ket[1]{{|{#1}\rangle}}
\usepackage{graphicx}
\usepackage{bm}
\usepackage[dvipsnames]{xcolor}

\usepackage{natbib}
\setcitestyle{square,numbers}
\usepackage[colorlinks]{hyperref}
\usepackage[dvipsnames]{xcolor}
\usepackage[authormarkuptext=name,commentmarkup=uwave]{changes}
\definechangesauthor[name={J\AA L}, color=green!70!black]{ja}
\definechangesauthor[name={KS}, color=blue]{ks}
\definechangesauthor[name={MRK}, color=red]{MRK}

\setcitestyle{square,numbers}
\usepackage[colorlinks]{hyperref}

\makeatother

\usepackage[most]{tcolorbox}
\newtcolorbox[auto counter]{problem}[1][]{%
    enhanced,
    breakable,
    colback=white,
    colbacktitle=white,
    coltitle=black,
    fonttitle=\bfseries,
    boxrule=.6pt,
    titlerule=.2pt,
    toptitle=3pt,
    bottomtitle=3pt,
    title={Important fact}~\thetcbcounter,
    #1}

\begin{document}
\title{Multipartite Bell-Greenberger-Horne-Zeilinger nonclassicality from interwoven frustrated down-conversion}

\author{Marek \.Zukowski}
\affiliation{International Centre for Theory of Quantum Technologies, University of Gdansk, 80-309 Gda{\'n}sk, Poland}

\author{Pawe\l{} Cie\'sli\'nski}
\affiliation{Centre for Quantum Technologies, National University of Singapore, Singapore 117543, Singapore}
\affiliation{International Centre for Theory of Quantum Technologies, University of Gdansk, 80-309 Gda{\'n}sk, Poland}
\affiliation{Institute of Theoretical Physics and Astrophysics, University of Gdańsk, 80-308 Gda\'nsk, Poland}
\affiliation{Faculty of Physics, Ludwig Maximilian University, 80799 Munich, Germany}
\author{Marcin Markiewicz}
\affiliation{International Centre for Theory of Quantum Technologies, University of Gdansk, 80-309 Gda{\'n}sk, Poland}
\affiliation{Institute of Theoretical and Applied Informatics, Polish Academy of Sciences, Ba{\l}tycka 5, 44-100 Gliwice, Poland}
\author{Konrad Schlichtholz}

\affiliation{International Centre for Theory of Quantum Technologies, University of Gdansk, 80-309 Gda{\'n}sk, Poland}

\begin{abstract} 

We present a theory of an interference process that starts with $N$ coherently pumped two-mode parametric down-conversion (PDC) sources, whose output modes are directed to $N$ observers such that each observer receives modes from two different source crystals. Each observation station is equipped with a locally controlled PDC crystal, coherently pumped with the source crystals, whose output modes are perfectly aligned with the input modes from the source PDCs. By varying the local phases of the input modes, perfect $2N$-photon interference is observed in $2N$ single-photon registrations, one in each output mode of the local PDCs. The interference results from the indistinguishability of the origins of the detected photons: either they all originate from the source PDCs or from the local PDCs. Bell–GHZ nonclassicality of the process emerges when one considers situations in which at least one of the local PDC processes is blocked and the $2N$-photon interference disappears. A ``lifted'' Clauser–Horne inequality is violated when its sole negative correlation term is tuned to maximal destructive interference, while all other terms involve settings in which one of the local pumps is off.

\end{abstract}

\maketitle
\section{Introduction}
In the early 1990s, Mandel's group performed a two-crystal parametric down-conversion experiment in which they observed interference depending on the degree of path indistinguishability, effectively erasing the information about which crystal produced the photons \cite{Zou_1991}.  An interference was observed in the rate of detection events at a single detector.  Zeilinger's  group performed an experiment nicknamed \textit{frustrated down-conversion} \cite{Herzog_1994, HERZ-1995}, in which a quantum superposition of emissions originating from two separate parametric down-conversion processes gives rise to an observable interference in the two photon emission process.  The setup was aligned such that the signal and idler emission modes emerging from the first crystal are fed into the second crystal and overlap with its signal–idler modes. The interference depends on phase shifts in the pump field and signal and idler modes between the crystals. 

New extensions of such a configuration, which could be, after a topological transformation, turned into Bell tests of non-classicality were given in \cite{PhysRevLett.118.080401}, for  further generalizations see \cite{RevModPhys.94.025007}. 
A trailblazing  experiment testing the interwoven frustrated down-conversion was reported in \cite{Qian_2023} and  \cite{MA}. In Ref. \cite{MA} a violation of a Bell inequality was claimed, but this was  premature, see \cite{PRL-2026}, and \cite{Price_2025,Wojcik2025}. In \cite{PRL-2026} it was shown that
the $2N$ photon interference process observed in \cite{MA} does have an explicit local realistic model. The model is based on ideas from Refs \cite{Larsson1998, Franson99,ModelNJP}.
Nevertheless, it was also shown in \cite{PRL-2026} that in the experiment of \cite{MA} the interference visibility of the probability of detecting a four-photon coincidence at measuring stations of Alice and Bob, with one  photon click at each detector,  was high enough to violate a proper Clauser-Horne inequality \cite{CH}.  However this  was possible under a different Bell experiment scenario than the one considered in \cite{MA}:  following ideas of   \cite{Hardy94,SinglephotonNJP,schlichtholz2023singlephoton} the pump powers enabling the down-conversion processes in the crystals at the local measuring stations are to be  variable and treated in the Bell-experiment protocol as additional parameters defining the settings of local measurements.  
In the experiment of \cite{MA} the settings were defined solely by local phase shifts.

We show an extension of these ideas to an $N$ observer configuration of interwoven parametric down-conversions, {\it the topology of which is presented for the {\em tripartite} case in Fig. 1 and  its caption}. A related topology of mode connections between beamsplitters instead of down-conversion crystals was proposed for generation of optical GHZ-type correlations with passive optics plus postselection, in a scheme \cite{Blasiak19} inspired by \cite{Yurke92}.

\begin{figure} \label{fig-1}
    \centering
\includegraphics[width=0.98\linewidth]{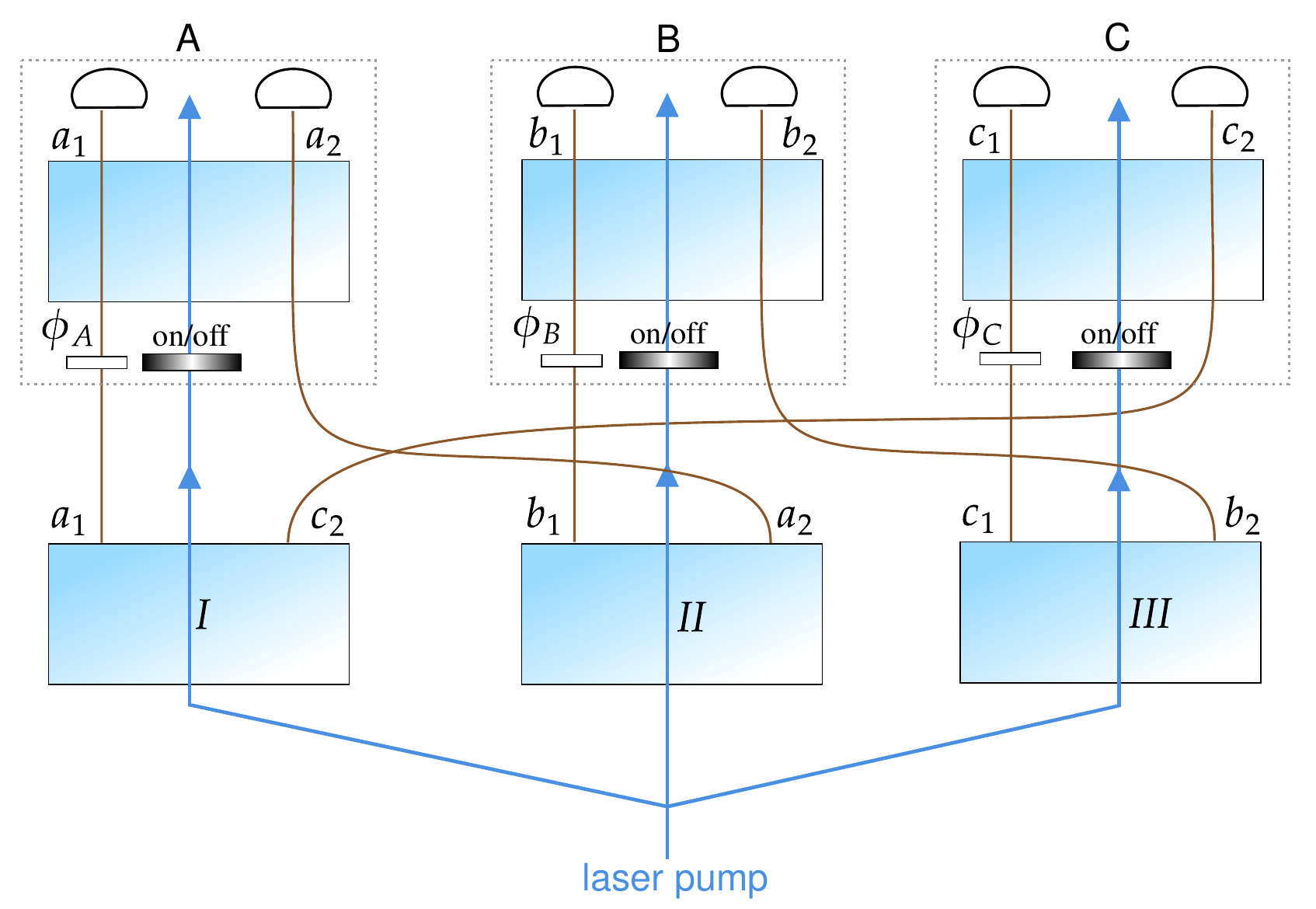}
    \caption{Three observer interwoven frustrated parametric down-conversion.
    Crystals I, II and III are a joint source of the initial state in modes $a_1c_2,b_1a_2, c_1b_2 $. Dotted boxes 
    represent the measuring stations A, B, C of Alice, Bob and Charlie. If the local pump reaching it is not blocked by the local  switch (marked as \textit{on/off}) the local crystal at a measurement station performs a unitary transformation on the respective local modes: $a_1a_2$, $b_1b_2$ and $c_1c_2$. These transformations depend on the local phase shifts $\phi_A, \phi_B, \phi_C$.
    To observe an interference in the probability of each of the six detectors to register single photons in coincidence all local pumps must be \textit{on}. To make this interference possible  three pair emissions from crystals I, II, III  have to be indistinguishable, at the positions of the detectors, from three pair emissions from A, B, C. To have a phase stability of the interference all pumps must be derived from a single laser, by beam-splitting the laser beam. The lengths of all optical paths shown in the figure, both of the pumping field(s) and all the signal and idler modes can differ only by much less than the coherence length of the laser pump. {\em An extension of the scheme to $N$ observers is obvious}.}
    \label{fig:placeholder}
\end{figure}

\section{The processes}
Fig. 1 shows three initial down-converters,  I, II, and III. The evolution of the initial vacuum state of all modes,  $\ket{\Omega}=\ket{0,0,0,0,0,0}$, can be well described here within the standard parametric approximation, in which the pumping field is represented by its amplitude (see, e.g., \cite{Pan12}). 
The dynamics is governed by the following Hamiltonian operators: first $H_{\textrm{I}}=\alpha a_1^\dagger c_2^\dagger+\overline\alpha a_1 c_2, $ second $H_{\textrm{II}}=\alpha b_1^\dagger a_2^\dagger+\overline\alpha b_1 a_2,$ third $H_{\textrm{III}}=\alpha c_1^\dagger b_2^\dagger+\overline\alpha c_1 b_2. $
These lead to unitary transformations described by $U_{\textrm{I}}(g)=\exp{i(ga_1^\dagger c_2^\dagger+\overline g a_1 c_2)}$, and similar formulas for $U_{\textrm{II}}$ and $U_{\textrm{III}}$, where $g=\alpha t$ and $t$
is the interaction time.

The down-converters at the measuring stations of Alice, Bob and Charlie (who sometimes will be denoted by  $O_ A, O_B$ and $O_ C$), if the pump powers are {\em on}, have Hamiltonians $H_{\textrm{A}}=\alpha a_1^\dagger a_2^\dagger+\overline\alpha a_1 a_2,$ and $H_{\textrm{B}}=\alpha b_1^\dagger b_2^\dagger+\overline\alpha b_1 b_2,$ and $H_{\textrm{C}}=\alpha c_1^\dagger c_2^\dagger+\overline\alpha c_1 c_2.$
The respective unitary transformations have the same relation with the Hamiltonians as the one addressed above, and will be denoted as $U_{\textrm X}(g)$, with X=A,B,C.

The phase-shift transformations in modes $a_1,b_1,c_1$ will be denoted by $U^{ph}_{\textrm X}(\phi_{X})$. 
If all down-converters at the measuring stations are \textit{on}, then the final states before measurement depend on the phases and form the following family 
\begin{equation}\label{FINAL}
    \prod_{\textrm{X=A,B,C}} [U_{\textrm{X}}(g)U^{ph}_{\textrm X}(\phi_{X})]\prod_{\textrm{K=I,II,III}}{U_{\textrm{K}}}\ket{\Omega}.
\end{equation}
Note that the initial state for the Bell-GHZ experiment is $\prod_{\textrm{K=I,II,III}}{U_{\textrm K}}\ket{\Omega}$, see Figure 1, and compare with \cite{PRL-2026}.

\section{Bell theorem for the configuration}
\subsection{Tripartite Clauser-Horne inequality}
We consider a lifted Clauser–Horne Bell inequality for the tripartite scenario,
\begin{align} \label{CH-INEQ} P(A,B,C') + P(A,B',C') + P(A',B,C') \nonumber \\ \quad - P(A',B',C') - P(A,C') - P(B,C') \le 0 . \end{align}
The primes indicate that the pump power at the local measuring station’s PDC crystal is switched on. All local events correspond to the detection of two photons at the given measuring station, as in \cite{PRL-2026}. For example, $P(A)=P(11|\textrm{\textit{off}},\phi_A)$ and $P(A')=P(11|\textrm{\textit{on}},\phi_A)$, where $11$ denotes the photon numbers detected in modes $a_1a_2$. 

Inequality (\ref{CH-INEQ}) can be derived from the generalization of the CHSH Bell inequalities for $N$ spatially separated systems, involving dichotomic observables, and two settings allowed for each of the $N$ parties. The complete family of such tight multipartite Bell inequalities, Werner-Wolf-Weinfurter-\.Zukowski-Brukner (WWWZB) ones, can be found in \cite{PhysRevA.64.032112-WW,Weinfurter2001101021, PhysRevA.64.032112-WW, Zukowski20022104011}.
For each $N$ we have $2^{2^N}$ of such inequalities, which wind up to an elegant single non-linear inequality. In \cite{PhysRevA.64.032112-WW} an explicit form for such inequalities for $N=3$ was given. These inequalities can be divided into
five equivalence classes with respect to permutations of the 
observers and the swapping of the labels of their measurement settings.
The form \eqref{CH-INEQ}  corresponds to a Clauser–Horne representation of one of the classes. The standard mapping between CHSH-type inequalities and their Clauser–Horne counterparts was used:
 by replacing $P(+|x,\lambda)-P(-|x,\lambda)$ with $2P(+|x,\lambda)-1$, where $P(\pm|x, \lambda) $ is the probability of the local hidden variable $\lambda$ to produce result $\pm1$, when the setting is $x$. This is possible when the local results are dichotomic, that is $P(+|x,\lambda)+P(-|x,\lambda)=1$. 
In the studied case local $+1$ result is the detection event denoted earlier as  $11$, all other local  events are treated as producing $-1$.

Inequality (\ref{CH-INEQ})  is a natural tripartite extension (``lifting'', \cite{Pironio05}) of the Clauser–Horne inequality \cite{CH}. Using the Bayes Rule, (\ref{CH-INEQ}) can be equivalently written as a  Clauser–Horne inequality for probabilities of the events for Alice and Bob, all conditioned on event $C'$ for Charlie.

  \subsection{Intuitive argument showing that the Bell inequality is violated}
To stress the importance of the source-(in)distinguishability and path identity in the processes, our reasoning will follow the approach based on the analysis of probability amplitudes of quantum processes presented by Feynman \cite{feynman1965}.
We treat all processes with amplitudes of order higher than $|g|^3$ as negligible. 
For $|g|^2<0.01$, the usual condition for achieving high visibility of interference in down-conversion experiments (see, e.g., \cite{Pan12}), this approximation introduces only small errors. 
In the  Appendix \ref{sec:appB} we provide analytic formulas for the probabilities entering the CH inequality. The probabilities do not contain terms of order $|g|^8$, and corroborate with  the intuitive approach.

(i) {\em Probability $P(A',B',C')$}: 
If PDCs A, B, C are \textit{on}, Fig.~1 shows that if all six detectors fire, each registering a single photon, the event may originate either from the simultaneous emission of two photons by each of the crystals I, II, and III, or from pair emissions in the crystals A, B, and C. 
No other combination of three crystals firing can lead to single-photon detections in all six modes. 
These two possibilities are indistinguishable beyond crystals A, B, and C, because the mode alignment imposes path identity \cite{PhysRevLett.118.080401, RevModPhys.94.025007}. 
The phase shifters located between the crystals introduce a controllable relative phase factor $e^{i(\phi_A+\phi_B+\phi_C)}$ for the amplitude of the process of three-pair emission from PDCs A, B, C. This we add to the amplitude of  the emission from I, II, III. 
For a negligible pump depletion, the overall probability amplitude is  $g^3(1+e^{i(\phi_A+\phi_B+\phi_C)})$, and the complete destructive interference is possible, as
\begin{equation}
    \label{PROBS-END} P(A',B',C') = 2 |g|^6 [1+\cos(\phi_A+\phi_B+\phi_C)].
\end{equation}

(ii) \textit{Probability $P(A,B,C')$}:
Since 11 events $A,B,C'$ involve three pair detections, one expects
$P(A,B,C')\propto |g|^6$,  as they require  emissions of three pairs. The 11 detection at Alice's station implies emissions from sources I and II to modes $a_1c_2$ and $b_1a_2$.
Thus, the photon detected by Charlie in mode $c_2$   originates from source I. Consequently his local PDC did not fire, and the photon in $c_1$ is from III.
Thus, all photons must originate from the source PDCs, and 
$P(A,B,C')=P(A,B,C)=|g|^6$. The phase $\phi_C$ is irrelevant.  

(iii) A similar reasoning shows that $P(A,B',C')=P(A',B,C')=P(A',B',C)=|g|^6$: also in these cases all photons must be from the source PDCs.

(iv) \textit{Probabilities $P(A,C')$ and $P(B,C')$}:
 Since the pump at Alice’s station is switched off, her two single photons  detections must be due to sources I and II firing, each a single pair. They  feed modes $a_1c_2$ and $b_1a_2$. 
Thus, the detected photon in mode $c_2$ of Charlie’s station is from source I. 
Hence, the photon in mode $c_1$ must come from the source III, and thus  we have a single pair emission into $c_1b_2$, by source III. Thus, event $11$ at Alice's and Charlie station, must be accompanied by event $11$ at Bob's, and the photons detected there originate from source PDC's II and III. It does not matter whether his local PDC is \textit{on} or \textit{off}.
Thus all source PDCs fired simultaneously. This implies $P(A,C')=P(A,B,C)=|g|^6$.   Because of the symmetry of the  configuration $P(B,C')=P(A,C')$.

Thus,  the quantum value of the lifted CH inequality is
\begin{align}
&CH_Q = -|g|^6[1+2\cos(\phi_A+\phi_B+\phi_C)].&
\end{align}
It reaches the maximum $CH_Q^{max}=|g|^6$ for the fully destructive interference, $P(A',B',C')=0$.
The CH inequality can be violated by choosing phase settings such that
$P(A',B',C')<|g|^6$.

\subsection{Critical visibility}
If in an experiment, due to some imperfections like less than perfect alignment, we get less than perfect visibility $V$ of the interference, that is we have  $P(A',B',C')=2|g|^6(1+V\cos{(\phi_A+\phi_B+\phi_C)})$, then the above formulas imply that we must have $V>1/2$ to violate local realism.
The threshold $V$ to violate local realism may be further increased by any noise and detection inefficiency. However, these are standard problems of all Bell-GHZ experiments.

\subsection{Remarks about tripartite character}
The lifted CH inequality (\ref{CH-INEQ}) is  a tripartite inequality, even though it seems that  Charlie enters it somehow trivially: he uses only one local pump setting: \textit{on}. If Charlie uses the \textit{off} setting instead, the inequality is not violated, as it then does not contain the interference term (\ref{PROBS-END}), compare (iii). 

One gets a symmetric inequality by summing the inequalities with  liftings involving each of the parties: 
$
CH^{A'}+CH^{B'}+CH^{C'}\leq 0.
$
Here $CH^{X'}$ stands for lifting the CH expression for two observers who are not $O_X$  by adding a conjunction with the event $X'=A',B',C'$ for respectively  observer $O_X=O_A,O_B,O_C$ to all probabilities of the original CH inequality. Now each party has two settings.
Due to the  symmetry of the configuration all of the inequalities $CH^{X'}\leq 0$ are violated for the same phases, and thus so is  the symmetric inequality. 

The correlations reveal \textit{genuine $n$-partite ``nonlocality"} \cite{Svetlichny87}. In Ref. \cite{Curchod19} a condition for that is given as
\begin{equation} \label{GEN}
    CH^{A'}+CH^{B'}+CH^{C'}-P(A',B',C')>0,
\end{equation}
and we have the property because the maximal violation of each $CH^{X'}$  is for $P(A',B',C')=0$. However, the additional term increases the threshold  visibility to achieve  (\ref{GEN}) to  $V_{genuine}= 5/8$.

\subsection{Bell Theorem via linear programming}
The problem of existence of a local realistic model for specific quantum predictions, from the point of view of numerical methods, is equivalent to linear programming, see \cite{zukowski1999strengtheningbelltheoremconditions} and \cite{PhysRevLett.85.4418}. Linear programming is equivalent to testing all possible tight Bell-inequalities for the given Bell-GHZ scenario.  We tested the quantum predictions with a current development of this method, e.g. used in \cite{Cieliski2025efficiency,Cieslinski2024polygamy}.
For the \textit{on-off} scenario, optimized over all possible phases with fixed small $g$, the linear programming  yields the lifted CH inequality (\ref{CH-INEQ}) as the one responsible for revealing the non-existence of a local realistic description. 

Note that the WWWZB inequalities  were formed for Bell-GHZ correlation functions. They are for dichotomic $\pm1 $ results and two settings per each observer. There seems to be no proof that Clauser-Horne versions of WWWZB inequalities exhaust full set of tight generalized Clauser-Horne inequalities for $N$ parties, and two settings for each observation station. Thus, linear programming is 
a tool at least as strong as all WWWZB inequalities. Still, in the studied case it returns the same result.

We applied linear programming to test whether Bell non-classicality can be observed in the situation in which all crystals are pumped (with the same strength) and only phase values  $\phi_A,\phi_B$ and $\phi_C$ acting as measurement settings, compare \cite{MA}. In \cite{PRL-2026}, it was shown that in the bipartite case of \cite{MA}, all of the relevant probabilities in the physically justified regime of $g$ can be explained by an explicit local hidden variable model. Linear programming strongly suggests that the same  holds for the process examined here, however perhaps only for two alternative settings for each observer (including Charlie). 
We scanned $\phi_A, \phi_B, \phi_C \in [0, 2\pi]$ with a step of $0.1\pi$ for $|g| \le 0.1$. 

\subsection{Generalization: Violation of local realism  for four (or more) parties}
There appears to be a single sequential topology, unique up to permutations of the source crystals, that provides a natural extension of the experiment for more than three parties. In this configuration each source PDC feeds modes associated with two observing parties, and the corresponding connectivity graph therefore forms a polygon.

We shall use a doubly lifted CH inequality:
\begin{align}\label{ch4}
P(A,B,C',D') + P(A,B',C',D') + P(A',B,C',D') \nonumber \\
\quad - P(A',B',C',D') - P(A,C',D') - P(B,C',D') \le 0 .
\end{align}
The reasoning is analogous to the three parties case.

In the interwoven configuration let the modes at the measuring stations be denoted by $a_1a_2$ (Alice), $b_1b_2$ (Bob), $c_1c_2$ (Charlie), and $d_1d_2$ (Dorothy),
and the source PDCs emit into the modes $a_1d_2$ (I), $b_1a_2$ (II), $c_1b_2$ (III), and $d_1c_2$ (IV).

(a) Assume that the PDC A at Alice’s station is switched off, while C, D  are \textit{on} or \textit{off}, and we have detections $11$ in all three, while the local PDC B may be \textit{on} or \textit{off}, and we do not assume what happens there.   The photons registered by Alice must therefore originate from source PDCs I and II. Consequently, a photon is detected in mode $b_1$, and it is from source II, and photon in $d_2$ is from source I. This excludes firing of PDC D. Therefore photon in $d_1$ must be from IV, but this produces a photon for $c_2$, thus C is not responsible for the photon at $c_1$, which means that III must have fired, providing the photons at $c_1$ and $b_2$. Thus, all source photons fired, and at B we have event $11$. 

(b) If PDC B is \textit{off}, while A, C, D may be \textit{on} or \textit{off}, the detections at $b_1b_2$ imply emissions into $b_1a_2$ from II and $c_1b_2$ from III. Thus at stations A and C photons are registered in $a_2$ and $c_1$. Therefore the photon at $c_2$ is from IV, which supplied a photon to $d_1$, and thus the photon at $d_2$ must be from I, and this implies a single photon at $a_1$. Thus all source PDCs fired and event 11 was registered at A.
All that means that $P(A, B^{(')}, C^{(')}, D^{(')}) =P(A,B,C,D)=P(A,C',D' )=P(B,C',D')$, where $^{(')}$ denotes prime or no prime (\textit{on} or \textit{off}).
As a consequence, no phase-dependent interference can occur. Interference effects arise only when all local PDCs A, B, C, D are switched \textit{on}, because the path identities then render the origin of the detected photons indistinguishable and allow phase-dependent interference between emissions from the source and local PDCs.

In ineq. (\ref{ch4}) the only interference term is $P(A',B',C',D')$. Its minimal value is zero. All other terms take the value $|g|^8$. Thus, inequality \eqref{ch4} is violated in the \textit{on–off} scenario. The reasoning can be generalized to configurations with an arbitrary number $N$ of observers, because of the following:

{\em If at least one local PDC is off and every station records one photon in each of its two output modes, then all detected photons must originate from the source PDCs; hence the corresponding probability is phase-independent and its lowest order term is $|g|^{2N}$. Moreover, if all but one station register the $11$ event, and the local PDC
is off at one of them, then the remaining station -- about which nothing was
assumed -- must register $11$ as well.}

\subsubsection{Clauser–Horne formulation and the absence of postselection}
{Note that the operational Bell interferometric scheme studied here does not rely on postselection, nor is it intended to generate GHZ states. With photon-number-resolving detection, the local detection events whose probabilities enter the generalized Clauser–Horne inequalities are directly distinguishable from all other local outcomes. Thus, if Alice registers an 11 event in a given experimental run, this outcome is unambiguously identified at the time of detection and is retained in the data used to estimate the relevant probabilities entering the CH Bell inequality. The same applies to the 11 events registered by Bob and Charlie. Crucially, no experimental run is accepted or discarded on the basis of a posteriori comparison of the detection records communicated between the parties; the classification of each local event is determined directly by the local measurement itself.
}

\section{A GHZ-like ``paradox''}
Assume that all higher corrections to the probabilities beyond $|g|^6$ are absolutely negligible. Then the following GHZ/Hardy-like reasoning, \cite{GHZ.89,Hardy94}  is possible.

Deterministic local hidden variable model for $P(11,11,11|\alpha_A,\alpha_B,\alpha_C;\phi_A, \phi_B,\phi_C)$ must be specified as:
\begin{equation}
\begin{split}
P(11,11,11 \mid \alpha_A,\alpha_B,\alpha_C;\phi_A,\phi_B,\phi_C) \\
= \int d\lambda\, \rho(\lambda)
  \prod_{X=A,B,C} p_X(11 \mid \alpha_X,\phi_X;\lambda) .
\end{split}
\end{equation}
with the obvious meaning of the notation, and determinism is when $p_X(11|\alpha_X, \phi_X; \lambda)=X(\alpha_X, \phi_X; \lambda)=0,1.$
We denote the \textit{on/off} settings of the pump at the PDCs by putting $\alpha_X=\alpha,0$.

The reasoning presented earlier shows that if two of the local pumps are switched \textit{off}, for example at Alice’s and Bob’s stations, and both stations yield the outcome $11$, then the outcome at Charlie’s station must also be $11$, regardless of whether his pump is \textit{on} or \textit{off}. This implies that
$A(0,\phi_A;\lambda)=1$ and $B(0,\phi_B;\lambda)=1$ together determine that $C(\alpha_C,\phi_C;\lambda)=1$, that is, $C=1$ for the same value of $\lambda$ and for arbitrary values of $\alpha_C$ and $\phi_C$.

By symmetry, analogous implications hold for permutations of the three observers. For example,
\[
(B(0,\phi_B;\lambda)=1 \ \text{and}\ C(0,\phi_C;\lambda)=1) \Rightarrow A(\alpha_A,\phi_A;\lambda)=1 ,
\]
and similarly
\[
(C(0,\phi_C;\lambda)=1 \ \text{and}\ A(0,\phi_A;\lambda)=1) \Rightarrow B(\alpha_B,\phi_B;\lambda)=1 .
\]
These relations follow from the symmetry of the configuration, which allows permutations of Alice, Bob, and Charlie while requiring that the pump setting is zero for two of the stations.

Thus, we have the following relations:
\begin{eqnarray} \label{A}
&B(0,\phi_B;\lambda)C(0,\phi_C;\lambda)A(0,\phi_A;\lambda)=1&\\
   \label{B} &B(0,\phi_B;\lambda)C(\alpha_C,\phi_C;\lambda)A(0,\phi_A;\lambda)=1&\\
  \label{C}  &B(\alpha_B,\phi_B;\lambda)C(0,\phi_C;\lambda)A(0,\phi_A;\lambda)=1&\\
  \label{D}  &B(0,\phi_B;\lambda)C(0,\phi_C;\lambda)A(\alpha_A,\phi_A;\lambda)=1.& 
\end{eqnarray}
All of them must hold for all $\lambda$'s for which the first one holds, because the first one implies all other three for the specific $\lambda$.
By multiplying the last three equations \eqref{B}-\eqref{D}  side by side, as for all $\lambda$'s singled out by the first relation one has $X(0, \phi_X; \lambda)^2=1,$ 
we get:
\begin{eqnarray} \label{Z}
&B(\alpha_B,\phi_B;\lambda)C(\alpha_C,\phi_C;\lambda)A(\alpha_A,\phi_A;\lambda)=1.&  
\end{eqnarray}
Hence we have a kind of GHZ paradox, as:
\begin{align}
&B(0,\phi_B;\lambda)C(0,\phi_C;\lambda)A(0,\phi_A;\lambda)=1
& \nonumber \\
&\Rightarrow B(\alpha_B,\phi_B;\lambda)C(\alpha_C,\phi_C;\lambda)
A(\alpha_A,\phi_A;\lambda)=1 .&
\end{align}
This means that
\begin{equation}\label{PARADOX}
\begin{split}
P(11,11,11 \mid 0,0,0;\phi_A,\phi_B,\phi_C) \\
\le P(11,11,11 \mid \alpha,\alpha,\alpha;\phi_A,\phi_B,\phi_C).
\end{split}
\end{equation}
But the quantum predictions give $P(11,11,11|0,0,0;\phi_A, \phi_B,\phi_C)= |g|^6$, and for specific phases $P(11,11,11|\alpha,\alpha,\alpha;\phi_A, \phi_B,\phi_C)=0$. 

While the 11-type probabilities entering the CH inequality have vanishing
$|g|^{8}$ coefficients, events with photon numbers other than $11$ at one of
the stations occur with probabilities of order $|g|^{8}$, i.e.,
$\mathcal{O}(|g|^{2})$ relative to the leading terms. Thus, for certain
$\lambda$'s satisfying Eq.~\eqref{A}, Eqs.~\eqref{B}--\eqref{D}
may fail with probability $\mathcal{O}(|g|^{2})$.
Thus, with the permutation symmetries of the problem  the relation (\ref{Z}) holds for $\lambda$'s satisfying (\ref{A})  with probability $1-\mathcal{O}(|g|^2)$. 
 
This reduces sharpness of the statement (\ref{PARADOX}) by a similar factor. Thus, for local hidden variable models  $ P(11,11,11|\alpha,\alpha,\alpha;\phi_A, \phi_B,\phi_C)=0$ is  still definitely excluded for small enough $|g|^2$. A more rigorous reasoning can be made using the techniques of Larsson \cite{LARSSON--XX}.

\section{Concluding remarks}
A multipartite configuration of  Fig. 1, 
with local \textit{on/off} control of the PDC processes at the measurement stations, leads to a violation of a “lifted” CH  inequality.
Bell–GHZ nonclassicality of the process can be shown when taking into account situations in which at least one of the local PDC processes is blocked. 

For the first-guess scenario of local settings 
defined by phases of modes entering the local measurement stations, with all pumps 
\textit{on} 
no violation of any two-setting Bell inequality was found. 
The quantumness of the process is revealed when one allows variable local pump powers (here we study the \textit{on/off} local pumping, but more general schemes are possible).

In the measuring stations we have active quantum optical processes which change the photon occupation numbers of the optical modes leading to the detectors. Photon occupation numbers are not merely undefined, like e.g. in 
\cite{3rdPaper}, but they change while interacting with local measuring stations. The  
quantum optical processes in the measuring stations are not just a trivial amplification or depletion of the optical fields, but quantum interference processes. 
Note further, that the observation in \cite{PRL-2026} that the vacuum term in the initial state produced by the source PDCs is essential for the interwoven frustrated interference to occur, can be easily shown  to apply to the processes considered here.

These unusual properties of the studied processes 
may lead to new results in foundations of quantum physics and new applied quantum physics.

\section*{Acknowledgments}
 This work was supported by project FENG.02.01–IP.05-0006/23 of the International Research Agendas Programme
MAB FENG (2021-2027), Priority FENG.02, Measure 
FENG.02.01, with the support of Foundation for Polish Science. P.C. acknowledges the FNP START scholarship and 
support of the National Research Foundation, Singapore 
through the National Quantum Office, hosted in A*STAR,
under its Centre for Quantum Technologies Funding Initiative
(S24Q2d0009).

\bibliography{bibliography-FRU}
\clearpage
\appendix
\begin{widetext}
\section{The initial state after the phase transformations}
\label{sec:appA}

\noindent The initial state which results from the action of all three
source Hamiltonians is:
    \begin{eqnarray}
\ket{\mathit{off},\mathit{off},\mathit{off}} = U_{\textrm{I}}U_{\textrm{II}}U_{\textrm{III}}\ket{\Omega}.
    \end{eqnarray}
Mode occupation number labeling is $\ket{n(a_1),n(a_2),n(b_1),n(b_2),n(c_1),n(c_2)}$ and phase shifts are denoted as $\phi_A$, $\phi_B$, $\phi_C$ . The $|\mathit{off},\mathit{off},\mathit{off}\rangle$ state up to $g^4$   after the phase transformations is
\begin{eqnarray}\label{off}
     &&U_{{\phi_A}}U_{{\phi_B}}U_{{\phi_C}}U_{\textrm{I}}U_{\textrm{II}}U_{\textrm{III}}\ket{\Omega}=\left(\frac{11 |g|^{4}}{8} - \frac{3 |g|^2}{2} + 1\right) | 0,  0,  0,  0,  0,  0\rangle \nonumber\\
    &+& \left(\frac{13 g^{3}\overline{g} e^{i ({\phi_B}+ {\phi_C})} }{6} - g^{2} e^{i( {\phi_B}+{\phi_C})}\right) | 0,   1,   1,   1,   1,   0\rangle + \left(\frac{13 g^{3}\overline{g} e^{i ({\phi_A}+ {\phi_C})} }{6} - g^{2} e^{i ({\phi_A}+ {\phi_C})}\right) | 1,  0,  0,  1,  1,  1\rangle\nonumber\\
    &+&  \left(\frac{13 g^{3}\overline{g} e^{i ({\phi_A}+ {\phi_B})} }{6} - g^{2} e^{i ({\phi_A}+ {\phi_B})}\right) | 1,  1,  1,  0,  0,  1\nonumber\rangle\nonumber\\
    &+&\left(- \frac{11 i g^{2}\overline{g} e^{i {\phi_C}} }{6}
    + i g e^{i {\phi_C}}\right) | 0,  0,  0,  1,  1,  0\rangle + \left(- \frac{11 i g^{2}\overline{g} e^{i {\phi_B}} }{6} + i g e^{i {\phi_B}}\right) | 0,   1,  1,  0,  0,  0\rangle\nonumber \\
    &+& \left(- \frac{11 i g^{2}\overline{g} e^{i {\phi_A}} }{6} + i g e^{i {\phi_A}}\right) | 1,  0,  0,  0,  0,  1\rangle \nonumber\\
    &+& \left(\frac{13 g^{3}\overline{g} e^{2 i {\phi_C}} }{6} - g^{2} e^{2 i {\phi_C}}\right) | 0,  0,  0,  2,  2,  0\rangle + \left(\frac{13 g^{3}\overline{g} e^{2 i {\phi_B}} }{6} - g^{2} e^{2 i {\phi_B}}\right) | 0,  2,  2,  0,  0,  0\rangle\nonumber \\
    &+&  \left(\frac{13 g^{3}\overline{g} e^{2 i {\phi_A}} }{6} - g^{2} e^{2 i {\phi_A}}\right) | 2,  0,  0,  0,  0,  2\rangle \nonumber\\
    &-& i g^{3} e^{3 i {\phi_C}} | 0,  0,  0,  3,  3,  0\rangle - i g^{3} e^{3 i {\phi_B}} | 0,  3,  3, 0,  0,  0\rangle - i g^{3} e^{3 i {\phi_A}} | 3,  0,  0,  0,  0,  3\rangle \nonumber\\
    &-& i g^{3} e^{i ({\phi_B}+2 {\phi_C})} | 0,  1,  1,  2,  2,  0\rangle - i g^{3} e^{i(2  {\phi_B}+ {\phi_C})} | 0,  2,  2,  1,  1,  0\rangle  - i g^{3} e^{i ({\phi_A}+2 {\phi_C})} | 1,  0,  0,  2,  2,  1\rangle\nonumber \\
    &-& i g^{3} e^{ i(2 {\phi_A}+ {\phi_B})} | 2,   1,  1,  0,  0,  2\rangle - i g^{3} e^{ i(2 {\phi_A}+ {\phi_C})} | 2,   0,  0,  1,  1,  2\rangle - i g^{3} e^{i( {\phi_A}+2 {\phi_B})} | 1,   2,  2,  0,  0,  1\rangle \nonumber\\
    &+& g^{4} e^{4 i {\phi_C}} | 0, 0,  0,  4,  4,  0\rangle + g^{4} e^{4 i {\phi_A}} | 4,  0,  0,  0,  0,  4\rangle + g^{4} e^{4 i {\phi_B}} | 0,  4,  4,  0,  0,  0\rangle\nonumber\\
    &+& g^{4} e^{i( {\phi_B}+3 {\phi_C})} | 0,  1,  1,  3,  3,  0\rangle +  g^{4} e^{i(3  {\phi_B}+ {\phi_C})} | 0,  3,  3,  1,  1,  0\rangle +  g^{4} e^{i ({\phi_A}+3 {\phi_C})} | 1,  0,  0,  3,  3,  1\rangle \nonumber\\
    &+& g^{4} e^{ i(3 {\phi_A}+ {\phi_B})}| 3,  1, 1,  0,  0,  3\rangle + g^{4} e^{ i(3 {\phi_A}+ {\phi_C})} | 3,  0,  0,  1,  1,  3\rangle + g^{4} e^{i ({\phi_A}+3 {\phi_B})} | 1,  3,  3,  0,  0,  1\rangle\nonumber\\
    &+& g^{4} e^{i(2  {\phi_A}+ {\phi_B}+ {\phi_C})} | 2,  1,  1,  1,  1,  2\rangle  + g^{4} e^{i( {\phi_A}+2 {\phi_B}+ {\phi_C})} | 1,  2,  2,  1,  1,  1\rangle + g^{4} e^{i( {\phi_A}+ {\phi_B}+2 {\phi_C})} | 1,  1,  1,  2,  2,  1\rangle\nonumber\\
    &+& g^{4} e^{i(2  {\phi_A}+2 {\phi_B})} | 2,  2,  2,  0,  0,  2\rangle  + g^{4} e^{i (2 {\phi_A}+2 {\phi_C})} | 2,  0,  0,  2,  2,  2\rangle  + g^{4} e^{i(2 {\phi_B}+2 {\phi_C})} | 0,  2,  2,  2,  2,  0\rangle\nonumber\\
    &-& i g^{3} e^{i( {\phi_A}+ {\phi_B}+ {\phi_C})} | 1,  1,  1,  1,  1,  1\rangle. 
\end{eqnarray}
\end{widetext}

\section{Probabilities of single photon counts in both local detectors at one, two and three observation stations, with local pumping switched ${\mathit  off}$ or on }
\label{sec:appB}
The calculations were done up to the fourth order in $g$ expansion of the unitary transformations. The  $|\mathit{off},\mathit{off},\mathit{off}\rangle$ initial state is shown in (\ref{off}). The  probabilities for 11 events on each site for all measurement settings in the studied Bell scenario are:
\begin{eqnarray} \label{PROBS}
    &P(\mathit{off}) = |g|^4 - \frac{10}{3}|g|^6 &\nonumber\\&+ \frac{205}{36}|g|^8 \approx |g|^4- \frac{10}{3}|g|^6,&\\
   & P(\mathit{on}) =|g|^2-\frac{8}{3}|g|^4 -\frac{65}{9}|g|^6&\nonumber\\& +\frac{278}{9}|g|^8 \approx |g|^2-\frac{8}{3} |g|^4 -\frac{65}{9}|g|^6,&\\
&P(A,B)=P(B,C)=P(A,C)=|g|^6,&\\
\label{1111}    &P(\mathit{off},\mathit{off},\mathit{off})&\nonumber\\&=P(\mathit{on},\mathit{off},\mathit{off})=P(\mathit{on},\mathit{on}, \mathit{off})= |g|^6,&\\
 \label{2222}   &P(\mathit{on}, \mathit{off})=|g|^6,&\\
   & P(on,on)=|g|^4-\frac{16}{3}|g|^6&\nonumber\\&+\frac{361}{36}|g|^8\approx|g|^4-\frac{16}{3}|g|^6,&\\
\label{PROBS-END-SUP}   &P(A',B',C')=2 |g|^6 [1+\cos(\phi_A+\phi_B+\phi_C)].&
\end{eqnarray}
Above we have used the following conventions:
\begin{eqnarray} \label{PROBS-3}
    &P({\mathit{off}})=P(A)=P(B)=P(C)&\nonumber\\
   &P(on) = P(A')=P(B')=P(C'),&\nonumber\\
    &P({\mathit{off}},{\mathit{off}},{\mathit{off}})=P(A,B,C)&\nonumber\\
    &P(on,{\mathit{off}},{\mathit{off}})&\nonumber\\&=P(A,B',C)=P(A,B,C')=P(A',B, C)&\nonumber\\
    &P(on,on,{\mathit{off}})&\nonumber\\&=P(A',B',C)=P(A,B',C')=P(A',B, C'),&\nonumber\\
    &P(on,{\mathit{off}})=P(A,B')=P(B,C')=P(A',C)&\nonumber\\&=P(A',B)=P(B',C)=P(A,C'),&\nonumber\\
&P(on,on)&\nonumber\\&=P(A',B')=P(B',C')=P(A',C').& 
\end{eqnarray}
These values were used in our linear programming computations. They agree with the values derived in the main text via the ''intuitive approach``.

\end{document}